\documentclass[aps,prl,twocolumn,superscriptaddress]{revtex4}
\usepackage{subfigure}
\usepackage{graphicx}
\usepackage{rotating} \usepackage{color}
 \usepackage{amsmath,amsthm}
 \usepackage{soul}
\usepackage{epstopdf}
\begin{document}

\preprint{1.0}

\title{Effects of macromolecular crowding on the collapse of biopolymers}


\author{Hongsuk Kang}
\affiliation{Chemical Physics and Biophysics Program, Institute of Physical Science and Technology, University of Maryland, College Park 20742}

\author{Philip A. Pincus}
\affiliation{Materials and Physics Departments, University of California, Santa Barbara, Santa Barbara, California 93106}

\author{Changbong Hyeon}
\affiliation{Korea Institute for Advanced Study, Seoul 130-722, Korea}

\author{D. Thirumalai}
\affiliation{Chemical Physics and Biophysics Program, Institute of Physical Science and Technology, University of Maryland, College Park 20742}



\begin{abstract}
Experiments show that macromolecular crowding modestly reduces the size of intrinsically disordered proteins (IDPs) even at volume fraction ($\phi$) similar to that in the cytosol whereas DNA undergoes a coil-to-globule transition at very small $\phi$. We show using a combination of scaling arguments and simulations that the polymer size $\overline{R}_g(\phi)$ depends on $x = \overline{R}_g(0)/D$ where $D$ is the $\phi$-dependent distance between the crowders. If $x\lesssim \mathcal{O}(1)$, there is only a small decrease in $\overline{R}_g(\phi)$ as $\phi$ increases. When $x\gg \mathcal{O}(1)$, a cooperative coil-to-globule transition is induced. Our theory quantitatively explains a number of experiments.  
\end{abstract}


\maketitle

The importance of crowding in biology is being increasingly appreciated because of the realization that cellular processes occur in a dense medium containing polydisperse mixture of macromolecules. 
A number of studies have been performed to understand  the role crowding particles play in inducing structural transitions in disordered chiral homopolymers \cite{Snir2005Science,Kudlay09PRL}, in protein \cite{Zhou08ARB,Elcock10COSB,Cheung05PNAS} and RNA folding \cite{Pincus08JACS,Kilburn10JACS,Denesyuk11JACS}, gene regulation through DNA looping \cite{Li09NaturePhysics},  genome compaction \cite{Kim11PRL}.
Some of the consequences of crowding  can be qualitatively explained using depletion interaction introduced by Asakura and Oosawa (AO) \cite{Asakura58JPS}. In the AO picture, the crowding particles, treated as hard objects, vacate the interstitial space in the interior of the macromolecule to maximize their entropy. 
As a result, an osmotic pressure due to crowders reduces the size of the macromolecule. 

The predictions based on the AO theory rationalize the impact of crowding effects on synthetic and biological polymers  qualitatively provided only excluded volume interactions between the crowding particles and the macromolecules  dominate. Even in this limit two questions of particular importance for experiments on biopolymers 
require scrutiny.  (i) What is the extent of crowding-induced compaction in finite-sized polymer coils? These systems are minimal models for unfolded and intrinsically disordered proteins (IDPs), and in some limits (random loop model) also provide a useful caricature of chromosome folding. (ii) For polymers with $N$ monomers, what is the dependence of the average radius of gyration, $\overline{R}_g(\phi)$ ($\equiv\langle R_g^2(\phi)\rangle^{1/2}$), as a function of the volume fraction $\phi$ and size of the crowders? It is important to answer these questions quantitatively to resolve seemingly contradictory conclusions reached in recent experiments.

Here, we answer these questions  using a combination of scaling arguments and computer simulations. The two  length scales  that determine the degree of polymer compaction in solution, with crowding particles interacting with each other and  the polymer via hard repulsions, are $\overline{R}_g(0)$ (the size of the coil at $\phi = 0$), and the average distance $D$ between the crowders.  
We propose a scaling relation to predict the dependence of $\overline{R}_g(\phi)$ on $\phi$ based on the expectation that when $D\lesssim\overline{R}_g(0)$  the osmotic pressure acting on the  polymeric chain should reduce the polymer size. 
If correlations between the crowding particles are negligible, as explicitly shown here using simulations for $\phi$ as large as 0.4, the maximum $\phi$ in the cytosol, then a scaling ansatz would suggest, 
$\overline{R}_g(\phi)=\overline{R}_g(0)f(x)$ where $f(x)$ is a function of the dimensionless variable $x=\overline{R}_g(0)/D$. 
For a given $\phi$, $D \approx (4\pi/3)^{1/3}\sigma_c\phi^{-1/3}$ where $\sigma_c$ is the radius of a spherical crowding particle, and
thus $x=(3/4\pi)^{1/3}\lambda\phi^{1/3}$, where $\lambda\equiv \overline{R}_g(0)/\sigma_c$. The form of $f(x)$ is difficult to calculate because of correlations in the fluid-like crowding particles \cite{Castelnovo04Macromol}. Nevertheless, we anticipate distinct scenarios in two limits of $x$. 
(i) When $x\sim\mathcal{O}(1)$, ($D\sim \overline{R}_g(0)$),  compaction of the coil should occur
without altering the chain statistics, $\overline{R}_g(\phi)=l_{\phi}N^{3/5}\sim \overline{R}_g(0)= l_0N^{3/5}$ where $l_0(l_{\phi})$ are the Kuhn lengths in the absence (presence) of crowding particles
; thus $f(x)\sim \mathcal{O}(1)$ implying that $\overline{R}_g(\phi)$ should depend weakly on $\phi$.  
(ii) In contrast, when $x\gg \mathcal{O}(1)$, ($D\ll\overline{R}_g(0)$), 
we expect that osmotic pressure 
induces collapse of the
polymer coil to a globule so that $\overline{R}_g(\phi)\sim N^{1/3}$.
These arguments suggest that the value of $x$ controls the polymer size ($N\gg 1$) in a crowded environment where only  excluded volume interactions are relevant.

With the two scenarios, expressed in terms of  $\overline{R}_g(0)$ and $D$ as a guide, we performed Langevin simulations in explicitly modeled spherical crowding particles with varying sizes, $\sigma_c$,  and for a range of $\phi$. 
Despite considerable efforts to predict the effects of crowders on polymer size \cite{frisch1979JPS, naghizadeh1978PRL, Grosberg82Biopolymers,Vasilevskaya95JCP,Krotova2010PRL}, it is difficult to accurately include the crucial effects of multi-particle correlations among crowding particles or strong correlation of monomers in a polymer chain (stiff or flexible) using phenomenological analytic theories  \cite{SokolovPRL03,Castelnovo04Macromol,Diamont00Macromolecules,Diamont00PRE} or microscopic formalism \cite{Shaw91PRA}.
To this end, we used a bead-spring model for the polymer and soft-sphere potentials to model interactions between the explicitly modeled crowder and the beads on the polymer \cite{Supple}. 
Due to interactions among polymer segments and crowders, the effects of semi-flexibility and polyelectrolyte nature of the polymer are important on the local scale $\lesssim l_p$ (persistence length)  \cite{RevModPhys.79.943}. 
However, in the length scale of our interest ($\gg l_p$) the self-avoiding bead-spring model suffices to capture the global characteristics of DNA. 
In this model,
local interactions can be accommodated by renormalization of the strength of volume exclusion.
Indeed, such models have been used to glean insights into chromosome folding \cite{lieberman2009Science}. 
Two variations of the random coils, one for IDPs, and the other for DNA, are used to cover a range of $x$ values. 

\begin{figure}[t]
\begin{tabular}{c}
\includegraphics[width=1.0\columnwidth]{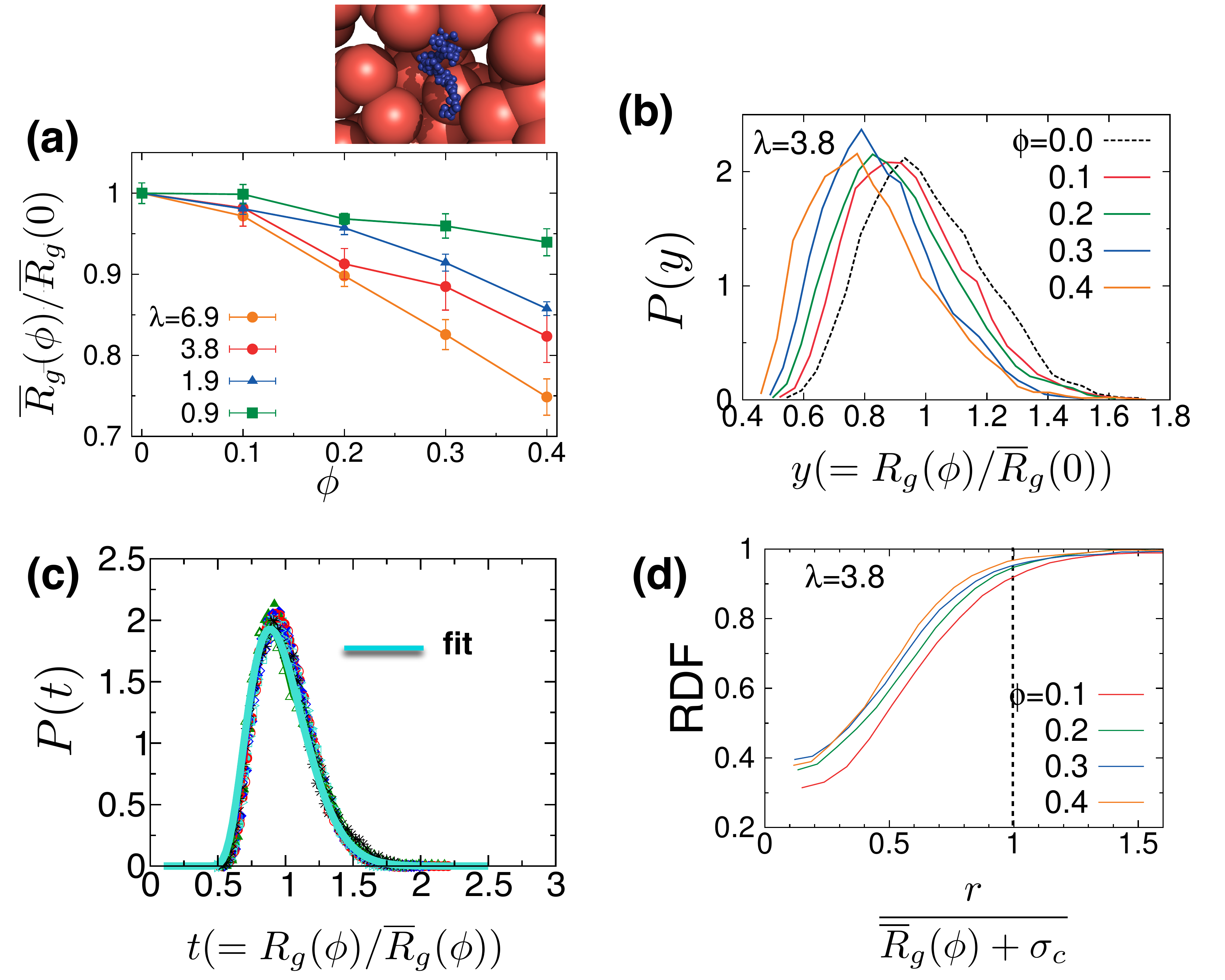}
\end{tabular}
\caption{
Crowding effect on the conformation of a SAW chain ($N=100$) for $\lambda \sim\mathcal{O}(1)$. 
(a) $\overline{R}_g(\phi)$ as a function of crowder volume fraction ($\phi$). 
The $\lambda$ values are shown in different colors. 
A snapshot of the SAW chain and crowding particles on the top is for $\lambda=1.9$ at $\phi=0.3$. 
(b) Distribution of $R_g$, $P(R_g(\phi))$, at $\lambda=3.8$. 
(c) Collapse of  $P(R_g(\phi))$ with $\phi=0.1-0.4$ and $\lambda=0.9-6.9$ onto a universal curve (Eq.\ref{eqn:PRg_scaling} with $b=1.120$ and $\mathcal{N}=13.69$) obtained by rescaling $R_g(\phi)$ by $\overline{R}_g(\phi) $ justifies that the statistics of the polymer coil does not change. 
The corresponding result for the end-to-end distance is in the SI.
(d) RDF of crowders from the center of the SAW chain at varying $\phi$. 
\label{Fig1}}
\end{figure}

{\it Compaction  due to large crowders $(\lambda \sim\mathcal{O}(1))$}:
If $D \sim \overline{R}_g(0)$ (scenario (i)) there ought to be only a modest reduction in the polymer size because the statistics of the polymer conformations (as assessed by distribution of $R_g(\phi)$, $P(R_g(\phi))$) are essentially unchanged. 
The reduction in $\overline{R}_g(\phi)$ 
becomes greater with increasing $\lambda$ (Fig.\ref{Fig1}a). 
However, the extent of compaction is only on the order of (5-8)\% for $\lambda<2.0$ (Fig.\ref{Fig1}a). 
At $\lambda=3.8$, as $\phi$ increases from 0 to 0.4, $P(R_g(\phi))$s clearly show a gradual shift towards smaller values of $R_g$ (Fig.\ref{Fig1}b). The $R_g(\phi)$ distributions plotted in terms of 
$t=R_g(\phi)/\overline{R}_g(\phi)$ for varying $\lambda$ values,  collapse onto a single universal curve (Fig.\ref{Fig1}c), 
corresponding to that of self-avoiding polymer \cite{Lhuillier88JP,deGennesbook}:
\begin{equation}
P(t)=\mathcal{N}e^{-(bt)^{-15/4}-(bt)^{5/2}},
\label{eqn:PRg_scaling}
\end{equation} 
where $b$ and $\mathcal{N}$ are parameters \cite{Supple}. 
The radial distribution functions (RDFs) of crowders from the center of polymer \cite{ramos2011macromolecular} (Fig.\ref{Fig1}d) show that the crowders are depleted from the space occupied by polymers.  
A snapshot from simulation (Fig.\ref{Fig1}a, top) shows that when $D\sim \overline{R}_g$ the polymer chain retains the shape with only modest compaction in the space between crowders. 

\begin{figure}[t]
\includegraphics[width=1.0\columnwidth]{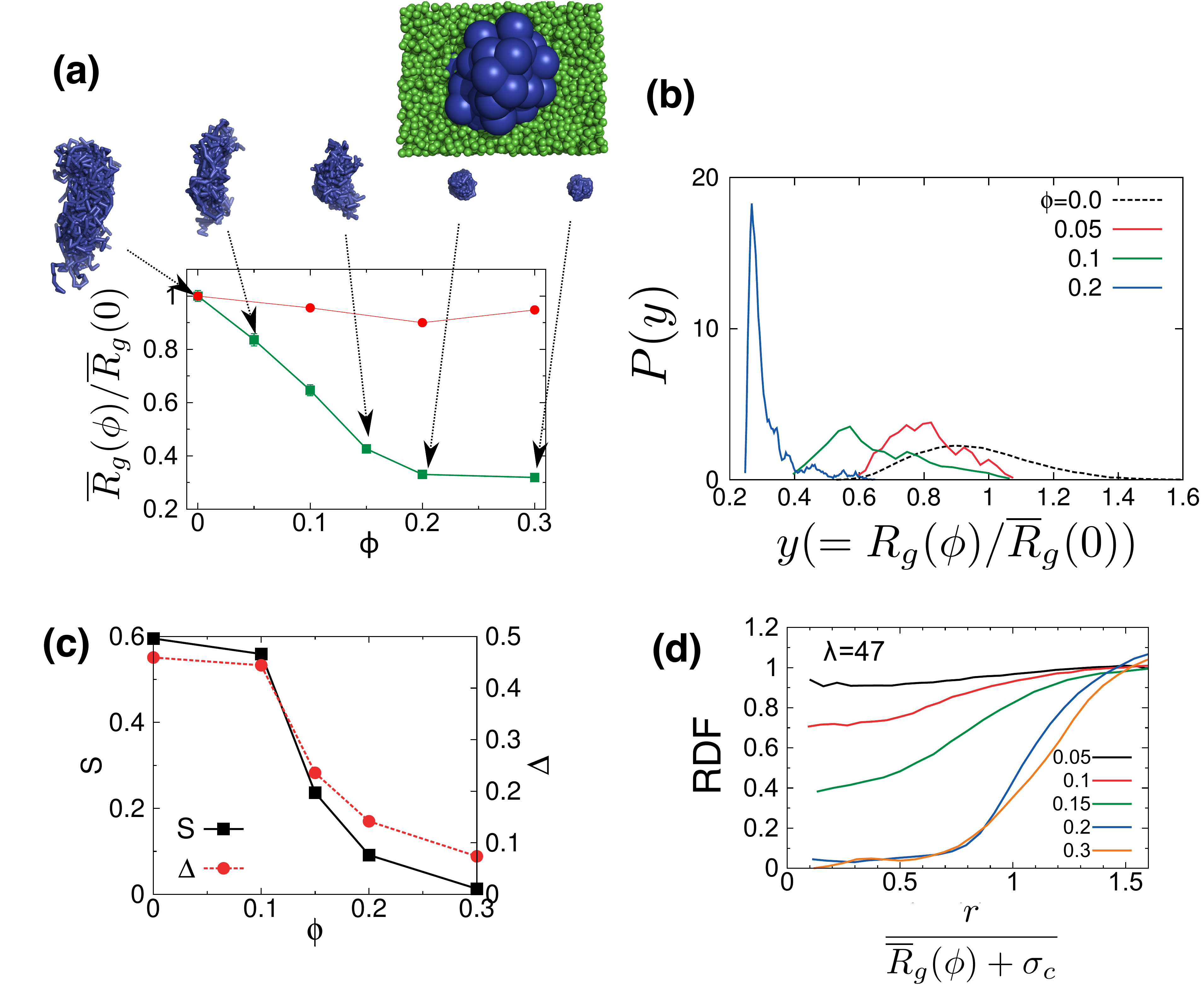}
\caption{Dramatic compaction of SAW chain ($N=50$) ($\lambda=47$). 
(a) $\overline{R}_g(\phi)/\overline{R}_g(0)$ of SAW chain 
(green square). 
The line with red circles is from Ref.\cite{Kim11PRL} that simulated the effect of macromolecular crowding on SAW polymer implicitly by using the effective depletion AO potential between two sites on a polymer.  
Ensemble of polymer conformations at each $\phi$ and a snapshot of simulation at $\lambda=47$ and $\phi=0.2$ are shown on the top.   
(b) $P(R_g(\phi))$ at $\lambda=3.8$. 
(c) Shape parameter ($S$) and asphericity ($\Delta$) of SAW chain as a function of $\phi$. 
(d) RDF of crowders from the center of position in the SAW chain at varying $\phi$.
\label{Fig2}}
\end{figure}

{\it Coil-globule transition due to small sized crowders $(\lambda\gg 1)$}:
When the size of the crowders is decreased there is a dramatic effect on the polymer size if $\phi>\phi_c$, where $\phi_c$ is a critical volume fraction for the coil-globule transition. 
Fig.~\ref{Fig2}a shows $\overline{R}_{g}(\phi)$ of a SAW polymer with $N=50$ for a minimal model of chromosome folding with crowders \cite{Kim11PRL}, which gives  $\lambda=47$. 
When $\phi$ increases to 0.3, $\overline{R}_g(\phi)$ reduces by 70 \% from the original size $\overline{R}_g(0)$.
The theoretical prediction,  $\overline{R}_g(\phi)/\overline{R}_g(0) \approx (1 - c\lambda \phi)^{\frac{1}{5}}$ \cite{Thirumalai88PRA} where $c$ is a constant, accounts for the simulation data in Fig.~\ref{Fig2}a for small $\phi\approx 0$. 
We note parenthetically that if the interaction between the beads were represented implicitly based on simulations of small $N$, as has been done previously \cite{Kim11PRL},  the dependence of $\overline{R}_g(\phi)$ on $\phi$ is qualitatively incorrect (red circles in Fig.~\ref{Fig2}a). 

The importance of the parameter $\lambda$ as a key determinant of the coil size can also be appreciated by comparing the results in Figs.~\ref{Fig1} \& \ref{Fig2}. 
In particular, $P(y)$ and RDFs with $\lambda=47$ (Fig.\ref{Fig2}) differ qualitatively from those with $\lambda=3.8$ (Fig.\ref{Fig1}). 
For $\lambda=47$, $P(R_g(\phi)/\overline{R}_g(0))$ with $\phi=0.2$ is peaked sharply at $R_g(\phi)/\overline{R}_g(0)\approx 0.25$ (Fig.\ref{Fig2}b) whereas at lower $\phi$ the peak is at $\approx 1$; and the rather abrupt coil-globule transition is striking given the small size of the polymer. 
RDFs for $\lambda = 3.8$ (Fig.\ref{Fig1}d) indicate that the crowders are essentially depleted from the region in which the polymer is localized for all $\phi$. In sharp contrast, for  $\lambda = 47$ (Fig.\ref{Fig2}d) at $\phi <  0.15$ there is a substantial probability that the crowders are in the vicinity of the polymer. 
Only after the coil-globule transition occurs at $\phi=\phi_c\approx 0.15 - 0.2$, the crowders are fully excluded from the interior of the polymer $r\lesssim \overline{R}_g$, and 
effectively no crowder particle is present in the interior of the collapsed polymer at $\phi\geq 0.2$ (Fig.\ref{Fig2}d).

To ascertain that the chain indeed forms a collapsed globule, we calculated the shape ($S$) and asphericity ($\Delta$) parameters 
\cite{NelsonJP86,Dima04JPCB}.
Both quantities, which measure the anisotropy of an object, are identically zero for a perfect sphere.
The ensembles of SAW configurations (Fig.\ref{Fig2}a) change from a prolate at low $\phi$ \cite{Solc71JCP,Honeycutt89JCP} to a spherical shape as $\phi$ increases.  The coil-globule transition, which has tricritical character \cite{Thirumalai88PRA,Duplantier88PRA},  is relatively sharp (Fig. \ref{Fig2}c) mirroring the decrease in $\overline{R}_g(\phi)$ (Fig.\ref{Fig2}a).      
At $\phi=0.3$, $S=0.01$ and $\Delta=0.07$ indicate that the polymer coil is collapsed to an almost perfect spherical globule.

{\it Critical $\phi$ for coil-globule transition.} 
The parameter $x$ is a useful measure for assessing whether a polymer of a given length in the presence of crowders of a specific size would undergo a coil-globule  transition. 
We estimate the critical $\phi$ ($\phi_c$) of crowders for a given parameter $\lambda$ by using  
$x_c = (3/4\pi)^{1/3}\lambda\phi_c^{1/3}$: 
\begin{equation}
\phi_c = \left(\frac{4\pi}{3}\right)\left(\frac{x_c}{\lambda}\right)^{3}.
\label{eqn:critical_phi}
\end{equation}
We estimate $x_c\approx 17$ because the polymer collapses at $\phi_c\approx 0.2$ for $\lambda=47$.    
The specific value of $x_c$ should in principle vary with the nature of interactions in the ternary system of polymer, crowding particles, solvent, and $N$. Nevertheless, the estimated $x_c$ is a guide to obtain an approximate estimate of $\phi_c$, and we show below it can be used to understand a number of experiments. Because of the restriction that $\phi_c < \phi_c^{max}\approx 0.74$ (close packing) and the weak dependence of $x$ on $\phi$ it follows from Eq.\ref{eqn:critical_phi} that as $\lambda$ decreases $\phi_c$ has to increase greatly in order for the crowding particles to induce coil-globule transition. Therefore, for small $\lambda$, one can only expect modest reduction in $\overline{R}_g(\phi)$ (Fig. 1a).


\begin{figure}[t]
\includegraphics[width=0.80\columnwidth]{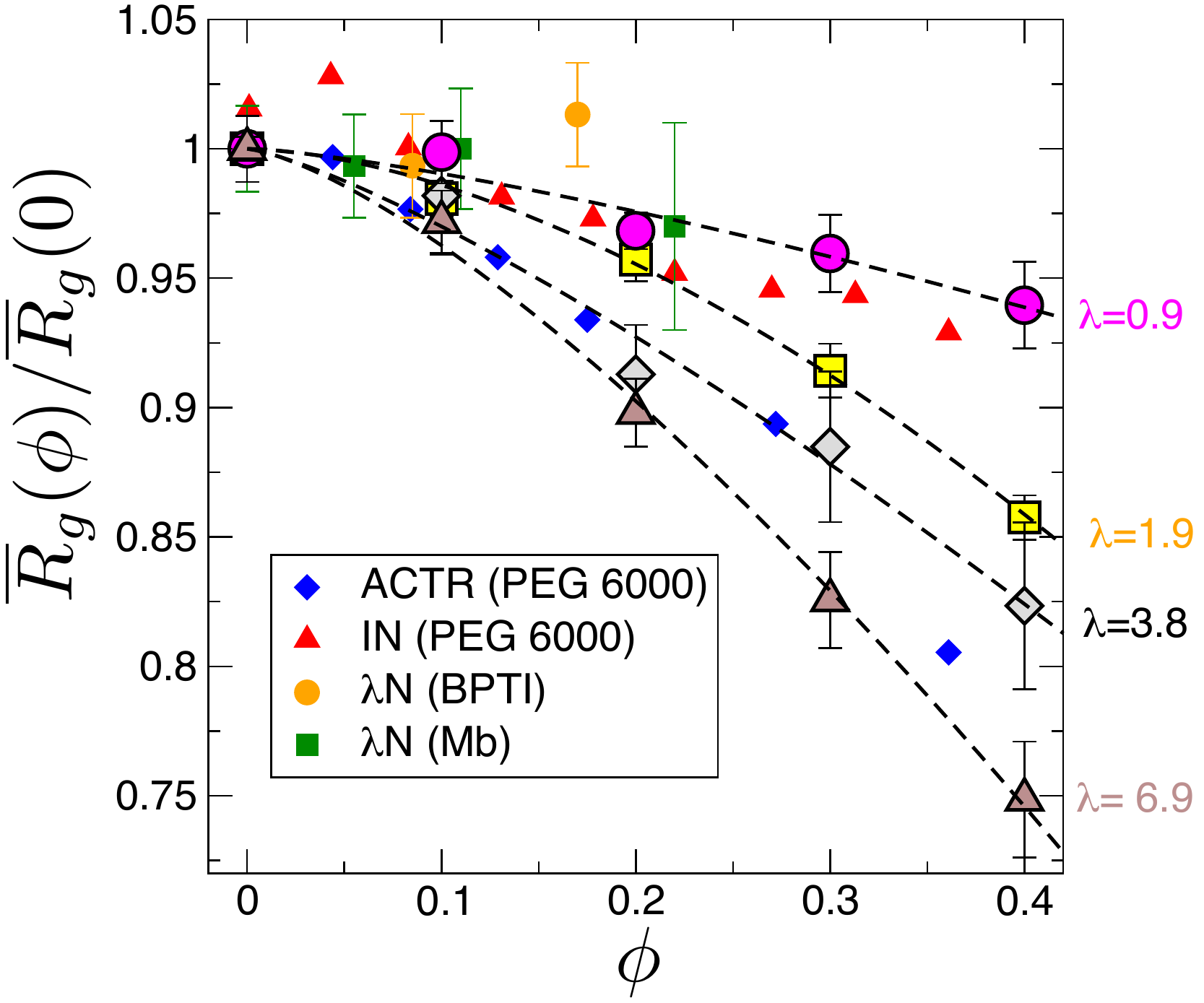}
\caption{Compaction of IDPs, ACTR and IN in an increasing volume fraction of PEG 6000 \cite{Soranno14PNAS}, and $\lambda$N in BPTI or metmyoglobin (Mb) \cite{Goldenberg14BJ}. 
To compare with  experiments, we superimposed our simulation results with $N=100$ and $\lambda=$ 0.9, 1.9, 3.8. 
The compaction of IN, $\lambda$N (BPTI), and $\lambda$N (Mb) is  described by $\lambda$=0.9, and ACTR  by $\lambda$=3.8.
\label{Fig3}}
\end{figure}

{\it Applications to experiments:} The combination of the scaling-type arguments 
and our simulation results offers a unifying framework for understanding experimental results on the effects of crowding on two entirely different classes of biopolymers.

(1){\it DNA:}
Since the discovery by Lerman \cite{Lerman71PNAS},
it has been noted that addition of polyethyleneglycol (PEG) to a coiled DNA induces cooperative coil-globule transition \cite{Vasilevskaya95JCP}.
Because $N \gg 1$ for DNA, collapse transition is accompanied by a substantial volume change ($N^{3\nu}\rightarrow N^1$). 
For T4-DNA whose contour length $L_c\approx 3.27\times 10^5$ nm and $l_p\approx 50$ nm \cite{Vasilevskaya95JCP}, 
$\overline{R}_g(0)\approx l_p (L_c/l_p)^{3/5}\approx 0.97\times 10^5$ nm, and $\sigma_c\approx 0.195\times P^{0.583}$ nm ($P$, polymerization index) for PEG \cite{Supple}, $\lambda\approx 4.97\times 10^5\times P^{-0.583}$, which leads to $\phi_c\ll 1$ for almost any $P$. 
Our theory shows that only a small amount of PEG is sufficient to induce coil-globule transition of DNA of a genomic size, as established experimentally.

(2){\it Intrinsically Disordered Proteins (IDP):}
There has been considerable interest in 
the effects of crowding on IDPs, which have critical functional roles especially in eukaryotes \cite{Babu12Science}. Based on recent single molecule  \cite{Soranno14PNAS} and small angle neutron scattering \cite{Goldenberg14BJ} experiments, it has been concluded that for certain IDPs crowding induces a very small ($\sim$ 5\%) reduction in the size whereas for others the effects are larger ($\sim$ 30\%).  
These results, which apparently cannot be explained by scaled particle theory that only accounts for excluded volume interactions, have lead to explanations that are difficult to rationalize \cite{Goldenberg14BJ}.  
Our theoretical results for neutral polymer coils nearly quantitatively account for the experimental findings for those IDPs with relatively small net charge per residue for which the polymer model used here is most appropriate. 
A typical IDP with $N\approx  100$ has $\overline{R}_g(0)\approx 3$ nm from 
$\overline{R}_g(0)\approx 0.193\times N^{0.598}$ \cite{Kohn04PNAS}. 
For IDP in the presence of PEG \cite{Soranno14PNAS}, we estimate 
$\lambda\approx 15.4\times  P^{-0.583}$, thus
$\phi_c\approx (4\pi/3)\times (x_c/15.4)^3\times P^{1.75}$. 
If $x_c$ is large, as is required for inducing globule formation, $\phi_c$ would be  greater than 
$\phi_c^{max}$ even for small PEGs with $P=1$.  
For PEG-6000 ($P\approx 136$) \cite{Soranno14PNAS}, $\lambda\approx 0.88$ and $\phi_c>\phi_c^{max}$. 
Therefore, the first conclusion is that there ought to be no coil-globule transitions in IDPs in \cite{Soranno14PNAS} using neutral crowders if one assumes IDP as a self-avoiding polymer.  
This is in accord with experiments probing crowding effects on five IDPs 
\cite{Soranno14PNAS,Goldenberg14BJ}.  

A more precise comparison with experiments can be made using our results for those IDPs with small net charge for which coil description is most appropriate. 
We consider the activator for thyroid hormones and retinoid
receptors, ACTR, and the N-terminal domain of the HIV-1 integrase,
IN 
with PEG as the crowding agent \cite{Soranno14PNAS}, and bacteriophage $\lambda$N with (nearly folded but likely hydrated)  BPTI and equine metmyoglobin  as crowding agents 
\cite{Goldenberg14BJ}. 
Fig.\ref{Fig3} shows $R_g(\phi)$ of IDPs.  
The excellent agreement between theory and experiments with {\it no adjustable parameters} shows that 
$\lambda$ 
controls the size. Thus, for these IDPs the present analysis, which relies on excluded volume as the dominant factor, suffices.   
Recently, other experimental studies have also noted that the size of unfolded proteins is insensitive to varying concentrations of crowders such as dextran, Ficoll, PVP, BSA, and lysozyme \cite{Goldenberg14BJ,Miklos11JACS,Wang12JACS}, leading the authors to suggest that attractive crowder-protein interactions, which compensates for the effects of excluded volume interactions, are at play.  
However, the $\lambda$ calculated for the systems in these studies all lie in the range of $\lambda \approx \mathcal{O}(1)$ and hence $x\sim1$, where the effect of neutral crowders on protein size is expected to be minimal.      
Thus, our  theory of neutral crowders based on two competing length scales, $\overline{R}_g(0)$ and $D$, fully explains a minimal effect of macromolecular crowding on IDPs in Refs. \cite{Goldenberg14BJ,Soranno14PNAS} and proteins in Ref. \cite{Miklos11JACS}. 

\begin{figure}[t]
\includegraphics[width=1.00\columnwidth]{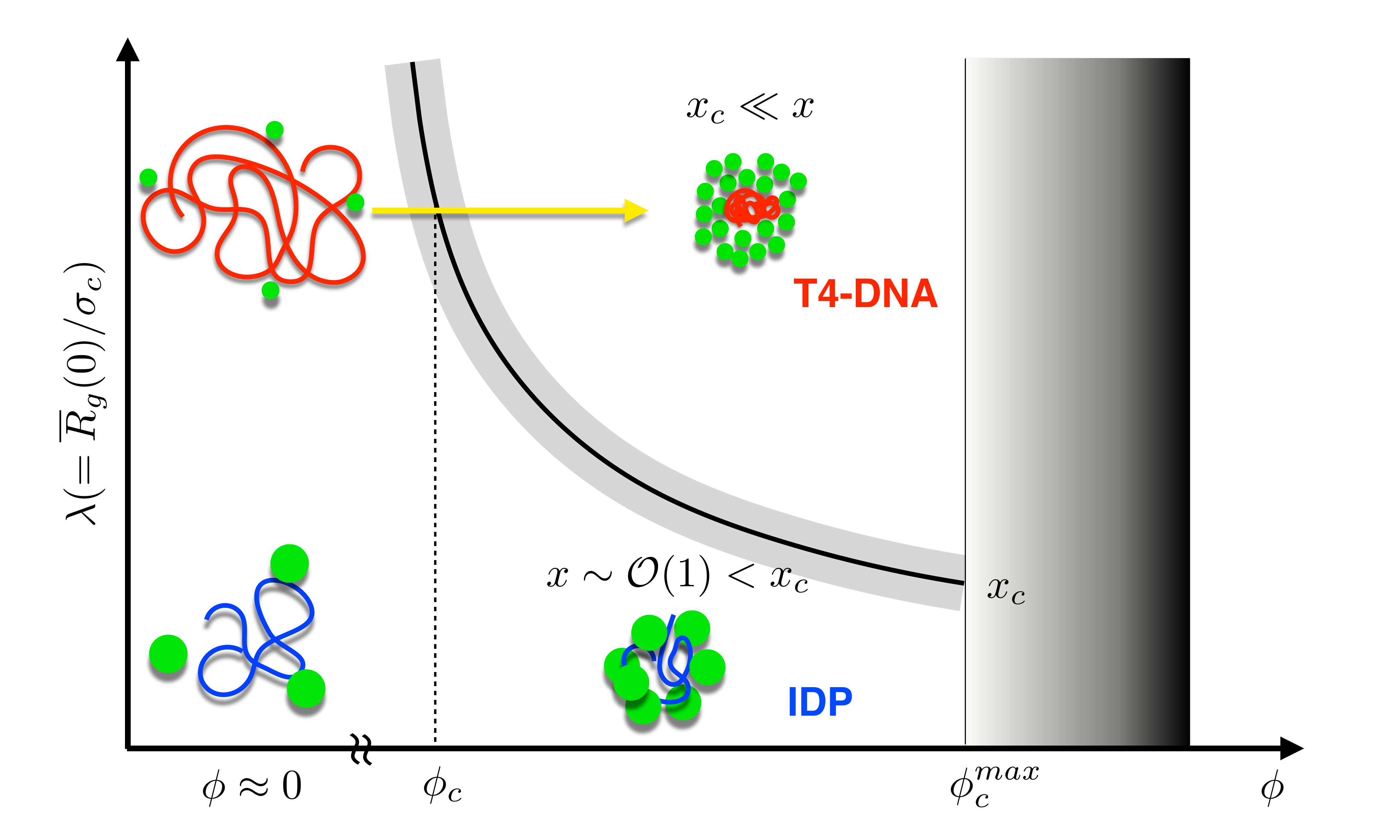}
\caption{Diagram of polymer collapse which varies depending on the value of parameter $x$.
For $x\sim\mathcal{O}(1)<x_c$, 
the size of polymer decreases with an increasing $\phi$ while maintaining its coil statistics ($\overline{R}_g\sim N^{3/5}$). By contrast, for $x_c\ll x$, the polymer undergoes coil-globule transition. 
$\phi_c$ greater than the volume fraction of close packing is not accessible ($\phi\gtrsim \phi_c^{max}$).  
Difference of the crowding induced dynamics in the two regimes of $x$ is illustrated with IDP and T4-DNA.    
\label{Fig4}}
\end{figure}

We conclude with a few additional remarks:
(i) It is tempting to use depletion potential obtained for small $N$ as a potential of mean force (PMF) for simulating a polymer with large $N$. 
 Kim \emph{et al.} \cite{Kim11PRL} obtained an effective $\phi$-dependent PMF between two beads of a small polymer in a crowded environment, 
and used the resulting PMF to simulate the crowding effect on the compaction of a long polymer.
As shown in Fig.\ref{Fig2}a (red circles), they found a slight non-monotonic turnover of $\overline{R}_g(\phi)$ at $\phi=0.2$, and ascribed their finding to the 
$\phi$-dependent repulsive barrier in the depletion potential. Their argument is that energy cost to squeeze out the crowders from the space between monomers increases with $\phi$. 
Our study, which simulates the SAW polymer in \emph{explicit} crowding particles with the identical parameters used in Ref.\cite{Kim11PRL}, shows a monotonic reduction of $\overline{R}_g(\phi)$ (green squares, Fig.\ref{Fig2}a).  
Crowding effects on chain conformations for long polymers 
require that the crowders  be explicitly treated at all scales. 
(ii) 
Based on the finding that the extent of crowding-induced polymer compaction or collapse is determined by the parameter $x$ $(=\overline{R}_g(0)/D)$
we propose a phase diagram (Fig.\ref{Fig4}), which should serve as a useful guide in anticipating the results of crowding experiments on biopolymers. 
Our estimate of $x$ explains that small amount of PEG suffices to induce coil-globule transition in DNA, 
whereas IDP whose size is $N\approx 100$ would not display collapse transition even at high $\phi$. 
(iii) For IDPs with highly charged residues the polymer model used here is inadequate because of electrostatic interactions as well as potential correlations between charged residues are not taken into account. It is likely that if a minimal polymer model for such systems is constructed, which will naturally involve additional length scale due to polyampholyte effects \cite{Barrat93EL,Ha97JPII} then scaling theories along the lines used here will provide insights into the effect of crowding particles.

\begin{acknowledgments}
We thank Shaon Chakrabarti, Mike Hinczewski, Himadri Samanta, and Pavel Zhuravlev for useful discussions. This work was supported in part by a grant from the National Science Foundation (CHE 13-61946).
\end{acknowledgments}


\clearpage 

\section{Supplemental Material}
\makeatletter 
\renewcommand{\thefigure}{S\@arabic\c@figure}
\makeatother 

{\bf Model:}
We used a bead-spring model for the flexible polymers.  A large spring constant is chosen so that the distance between two successive monomers remains approximately a  constant.   We chose Weeks-Chandler-Anderson (WCA) potential for excluded volume interactions between monomers, and soft-sphere potential for crowder-crowder and crowder-monomer. 

The energy function for the system consisting of the self-avoiding walk (polymer)  and soft spherical crowders is
\begin{equation}
H_{tot}=H_B+H_{m-m}+H_{m-c}+H_{c-c}.
\end{equation}
Here $H_B=K\sum_{i=1}^{N-1}\left(|\vec{r}_{i+1}-\vec{r}_{i}|-l_0\right)^2/l_0^2$ 
is the bond potential along the polymer chain,  $K$ is the spring constant with $l_0$ being the bond length.  
We used the WLC potentials  for soft-core repulsion, so that $H_{m-m}=\sum_{i<j}^{N}\epsilon\left[(\frac{\sigma_{ij}}{r_{ij}})^{12}-(\frac{\sigma_{ij}}{r_{ij}})^6\right]\Theta\left(\frac{\sigma_{ij}}{r_{ij}}-1\right)$ with $\Theta(\ldots)$ being a Heaviside step function.   
$\epsilon$ is Lennard-Jones energy constant controlling the strength of the excluded volume interaction, and $\sigma_{ij}$ is the distance between two particles in direct contact, given by $\sigma_{ij}=\sigma_i+\sigma_j$ with $\sigma_{i}$ being the radius of a bead, where $i$ and $j$ are either the index for monomers in polymer chain or for crowding particles. 
Lastly, $H_{m-c}=\sum_i^{N}\sum_{j}^{n_c}\epsilon\left(\frac{\sigma_{ij}}{r_{ij}}\right)^{12}$ and $H_{c-c}=\sum_{i<j}^{n_c}\epsilon\left(\frac{\sigma_{ij}}{r_{ij}}\right)^{12}$ are the monomer-crowder and crowder-crowder repulsions, respectively. 
We simulated for $N=$50, 100, and adjusted $n_c$ to achieve the range of crowder volume fraction ($\phi=n_c\times \frac{4}{3}\pi\sigma_c^3/V$, where $V=L_x\times L_y\times L_z$ is the volume of periodic box) between 0 and 0.4. In order to investigate the effect of the size of crowders, we used $N=100$ 
with $\sigma_{c}=3.6\sigma_{m}$.

In order to compare our results with the results obtained from simulations with implicit crowders \cite{Kim11PRL}, we chose the same parameters used in ref \cite{Kim11PRL}. 
In particular, the number of monomers is $N=50$ and $\sigma_{m}=5\sigma_{c}$. 
Other simulation parameters are listed in Table I. \\

{\bf Simulation details:} In order to obtain adequate sampling of the conformational space of the system, we performed low friction Langevin dynamics (LFLD) \cite{HoneycuttBP92}. 
We followed the same simulation procedure described in detail elsewhere 
\cite{VeitshansFoldDes97}. 
The data for analysis were collected after $2\times10^{5}$ time steps for equilibration. The number of crowders varies according to chain conformation with the average value being $\sim 10^{4}$ for $\phi=0.3$.\\

\begin{table}
\begin{ruledtabular}
\begin{tabular}{cccccccccc}
$K$&$l_{0}$&$k_{B} T$&$\Delta t$&$\zeta_{m}$&$\zeta_{c}$\\
\hline
$1500\epsilon$&1.11 $\sigma_{m}$&0.6 $\epsilon$ & 0.01 $\tau$ & $0.05 m \tau^{-1}$ & $\zeta_{m} \left( \frac{\sigma_{c}}{\sigma_{m}}\right)$
\end{tabular}
\end{ruledtabular}
\caption{\label{table:parameters}Parameters characterizing the model. Lennard Jones energy constant $\epsilon$, the diameter of monomer $\sigma_{m}$ and $\tau=\sqrt{\frac{m\sigma_{m}^2}{\epsilon}}$ are used as the unit energy, the unit length and the unit of time, respectively. $K$ is a spring constants for a chain connectivity between monomers, $l_{0}$ is a bond length between monomers of a chain, $k_{B}T$ is a temperature, $\Delta t$ is a simulation time step, $\zeta_{m}$ and $\zeta_c$ are the friction coefficients for monomer and crowding particles, respectively.
}
\end{table}

\begin{figure}[t]
\includegraphics[width=1.0\columnwidth]{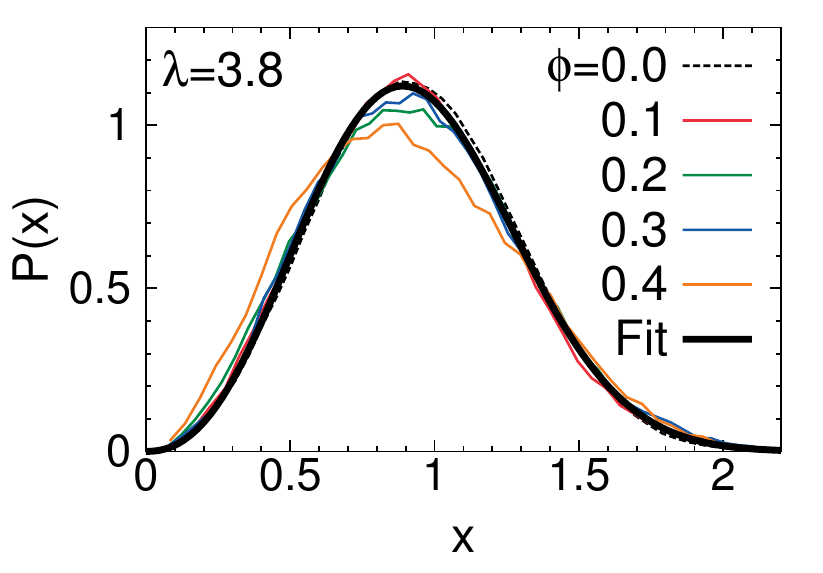}
\caption{The end-to-end distance ($R_{ee}$) distribution of a SAW scales as 
$P(x)\sim x^{g}$ at $x\rightarrow 0$ and $P(x)\sim e^{-x^{\delta}}$ at $x\gg 1$ where $x=R_{ee}(\phi)/\overline{R}_{ee}(\phi)$ with $\overline{R}_{ee}\sim N^{\nu}$ and $\delta=(1-\nu)^{-1}$ \cite{deGennesbook}.  
All the $P(R_{ee})$s with varying $\phi=0.0-0.4$ are collapsed onto $P(x)=\mathcal{N} x^{g+2}\exp{\left[-bx^{\delta}\right]}$ with 
$\mathcal{N}=3.68$, $g=0.30$, and $b=1.23$ with $\delta=2.5$ fixed. 
Note that $g\approx (\gamma-1)/\nu=0.283$ ($\nu=0.588$, $\gamma=7/6$) \cite{deGennesbook}. 
As the perturbation ($\phi$) increases, we observe an increasing deviation from $P(x)$ with $\phi=0$.  
\label{EtE}}
\end{figure}

{\bf Expression of $P(R_g)$ from scaling argument:} 
Based on the argument that the radius of gyration $R_g$ strongly decreases for $R_g<N^{1/d}$ and $R_g>N$ and that the distribution of the end-to-end distance $x$ decays as $P(x)\sim e^{-x^{\delta}}$ \cite{deGennesbook}, Lhuillier has proposed $P(R_g)$ of SAW in $d$ dimension as follows \cite{Lhuillier88JP}, which we slightly modified for our analysis, 
\begin{equation}
P_N(R_g)\sim \exp{\left\{-N\left[\left(\frac{N}{R_g^d}\right)^{\alpha}+\left(\frac{R_g}{N}\right)^{\delta}\right]\right\}}. 
\end{equation}
The first term in the exponent arises from the repulsion energy and the second term corresponds to elastic energy, which renders $P_N(R_g)\rightarrow 0$ when $R_g$ is either too small or too large. 
When $P(R_g)$ is written in terms of the dimensionless variable $t=R_g/\overline{R}_g=R_g/bN^{\nu}$, then 
$P(t)\sim \exp{\left[-(bt)^{-d\alpha}N^{1+\alpha(1-\nu d)}-(bt)^{\delta}N^{\delta(\nu-1)+1}\right]}$. 
Noting the property that $P(t)$ is independent of $N$, we can determine $\alpha=(\nu d-1)^{-1}=5/4$ and $\delta=(1-\nu)^{-1}=5/2$, which leads to Eq.1 in the main text.
\\

{\bf Size of PEG:}
In PEG-6000, 6000 corresponds to the polymer molecular weight ($M_w$ (g/mol)). 
The average radius of gyration of PEG with $M_w$ is  
$\sigma_c=0.0215\times M_w^{0.583}$ nm \cite{Linegar2010ColloidJournal}. 
Since molecular weight of PEG monomer is $\approx 44$ g/mol, we can convert the expression of $\sigma_c$ in terms of molecular weight into the one in terms of the degree of polymerization $P$: $\sigma_c\approx 0.195 \times P^{0.583}$ nm.

\end{document}